\documentclass[aps,prb,twocolumn,raggedbottom,floatfix]{revtex4}

\usepackage{amsmath,amsfonts,amssymb,amsthm,graphics}
\usepackage[next]{inputenc}
\usepackage[dvips]{epsfig}
\usepackage[colorlinks=true,citecolor=blue,linkcolor=blue]{hyperref}
\usepackage{bbm}
\usepackage{bbold}
\usepackage{booktabs}
\usepackage{multirow}
\usepackage{hhline}
\usepackage{amssymb}
\usepackage{comment}

\usepackage{bm}
\usepackage{color}
\usepackage{graphicx,latexsym}
\usepackage{epstopdf}

\usepackage{esvect}
\usepackage{fixmath}

\def\be{\begin{equation}}
\def\ee{\end{equation}}

\def\bi{\begin{itemize}}
\def\ei{\end{itemize}}
\def\bn{\begin{enumerate}}
\def\en{\end{enumerate}}
\def\bea{\begin{eqnarray}}
\def\eea{\end{eqnarray}}
\newcommand{\bpm}{\begin{pmatrix}}
\newcommand{\epm}{\end{pmatrix}}

\def\ba{\begin{array}}
\def\ea{\end{array}}
\def\bd{\begin{displaymath}}
\def\ed{\end{displaymath}}

\def\ra{\rangle}
\renewcommand{\imath}{\hspace{1pt}\mathrm{i}\hspace{1pt}}

\begin{document}
\title{Vortex bound states of charge and magnetic fluctuations-induced topological superconductors in heterostructures}
\author{Hossein Hosseinabadi}
\affiliation{Department of Physics, Sharif University of Technology, Tehran 14588-89694, Iran}
\author{Mehdi Kargarian}
\email{kargarian@physics.sharif.edu}
\affiliation{Department of Physics, Sharif University of Technology, Tehran 14588-89694, Iran}

\begin{abstract}
The helical electron states on the surface of topological insulators or elemental Bismuth become unstable toward superconducting pairing formation when coupled to the charge or magnetic fluctuations. The latter gives rise to pairing instability in chiral channels $d_{xy}\pm i d_{x^2-y^2}$, as has been observed recently in epitaxial Bi/Ni bilayer system at relatively high temperature, while the former favors a pairing with zero total angular momentum. Motivated by this observation we study the vortex bound states in these superconducting states. We consider a minimal model describing the superconductivity in the presence of a vortex in the superconducting order parameter. We show that zero-energy states appear in the spectrum of the vortex core for all pairing symmetries. Our findings may facilitate the observation of Majorana modes bounded to the vortices in heterostructures with no need for a proximity-induced superconductivity and relatively large value of $\Delta/E_F$.  
\end{abstract}
\date{\today}

\maketitle

\section{Introduction}
For many years it was a common belief that quantum mechanical particles are either bosons or fermions whose wave functions take a plus or minus sign, respectively, upon the exchange of two identical particles. But in the past forty years it has been realized that this picture for point-like particles is correct only for a space with dimensions equal or greater than three. There are other possibilities in two-dimensional (2D) space \cite{Wilczek:prl82}. In 2D systems, particles' wave functions can in general acquire a complex phase upon turning one particle around another one, the so-called Abelian anyons \cite{Leinaas1977,Wilczek:prl82'}. The first physical realization of Abelian anyons occurred in systems exhibiting the fractional quantum Hall effect \cite{Halperin:prl84,Arovas:prl84}. For the non-Abelian anyons, on the other hand, the multi-component wave functions live in a degenerate subspace \cite{Frohlich88}, and the interchange of particles amounts to a unitary evolution matrix within the degenerate subspace. The non-commutative structure of matrices promises a platform for fault-tolerant topological quantum computations \cite{Nayak:rev08}. 

Majorana fermions, a special class of non-Abelian anyons, discovered first by theoretical particle physicist Ettore Majorana \cite{Majorana37} in 1937, have the property that they are their own anti-particles. If $\gamma_i$ and $\gamma^\dagger_i$ are donated as the annihilation and creation operators for a Majorana fermion in a quantum state $|i\rangle$ then $\gamma_i=\gamma^\dagger_i$ and $\left\{ \gamma_i,\gamma_j \right\}=2\delta_{ij}$. Majorana fermions were not observed in elementary particle physics. But the developed concepts were traced in condensed matter physics years after the theoretical discovery \cite{Kopnin:prb91,Moore:npb91,Read:prb2000}. In the seminal work by Reed and Green \cite{Read:prb2000}, it is shown that a boundary between a 2D topological $p+ip$ superconducto and a trivial one hosts a single Majorana mode, and in a vortex core a Majorana bound state (MBS) is formed. A prime example of a system hosting MBS was introduced by Kitaev \cite{Kitaev2000}. The model is a chain of spin-less fermions with superconducting $p$-wave order parameter. Under certain conditions, where the bulk of the system is topologically non-trivial, two unpaired MBSs appear at the ends of an open chain. 

\begin{figure}[t!]
\begin{center}
\includegraphics[width=8cm]{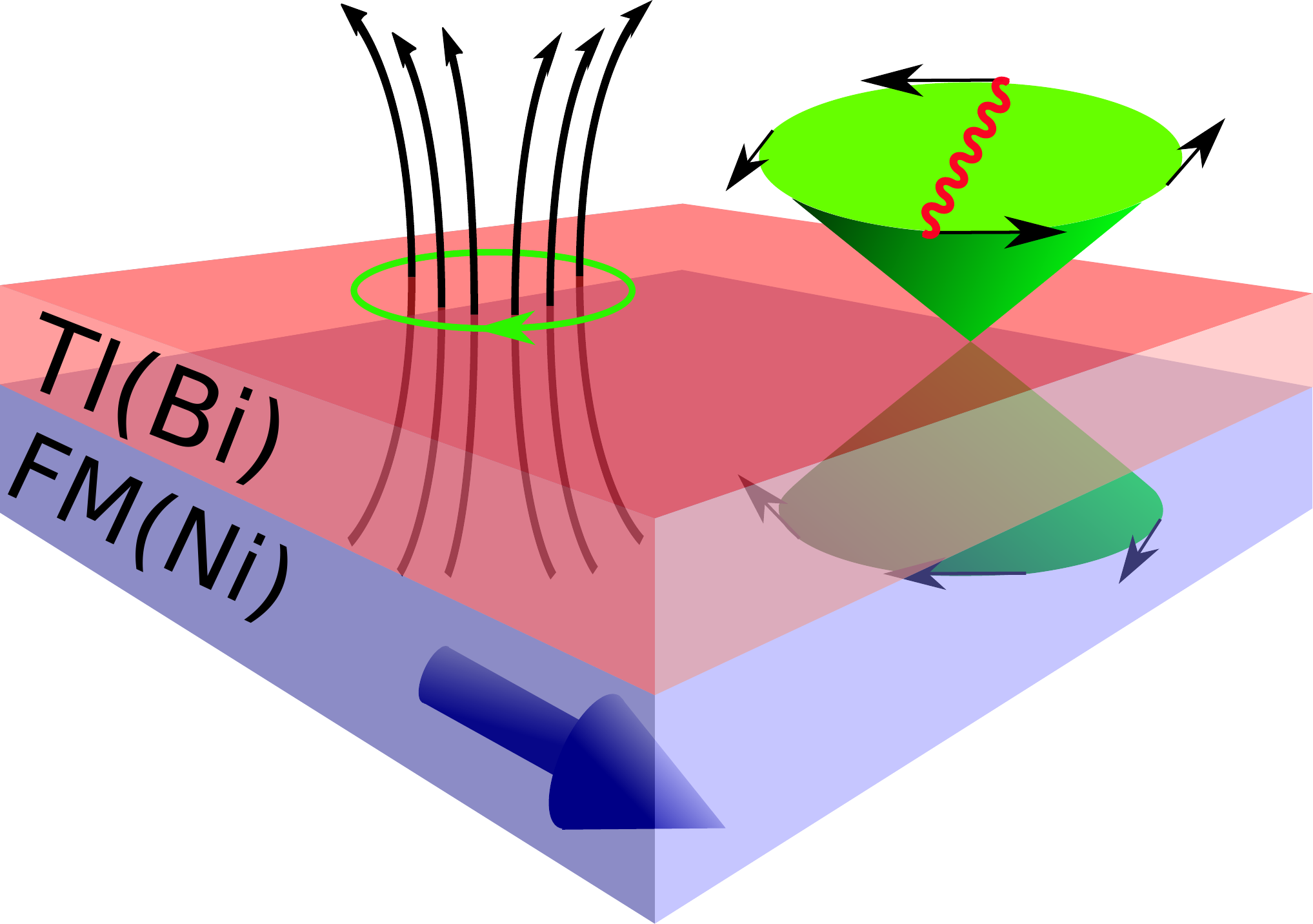}
\caption{A schematic representation of epitaxial bilayer. A thin film of Bismuth (Bi) or topological insulator (TI) is deposited on a ferromagnetic layer such as nickel (Ni), an experimentally realized bilayer \cite{Gong:sc17}. The Dirac cone and the black arrows indicate the helical electron states near the Fermi surface. The wavy red line shows the pairing states between electron states living son opposite sides of the Fermi surface, induced by the magnetic fluctuations on the in-plane magnetic moments (thick blue arrow) of nickel. Vertical arrows demonstrate a schematic view of a typical vortex.}\label{fig1}
\end{center}
\end{figure}

The discovery of topological insulators (TI) \cite{Kane:rmp10,Hasan:annalsrev11} provided a boost to the realization of Majorana fermions in solid state systems. In a celebrated work by Fu and Kane in Ref. [\onlinecite{Fu:prl08}] it is shown that the MBSs appear in the vortex cores of a conventional $s$-wave superconductor proximitized to the surface of TIs arising from the spin-momentum locked structure of the surface states of the TI and the underlying Berry phase, which makes the $s$-wave pairing formally like a $p$-wave pairing upon projection onto the surface states. Viewed the surface states as massive Dirac electrons, the vortex bound states of the corresponding fermion-vortex problem have been studied \cite{Jackiw:npb81,Seradjeh:npb08}. The MBSs can also be realized in doped TIs \cite{Vishwanath:prl11,Hughes:prl12}. Moreover, the surface of TIs can be replaced by more conventional semiconductors with strong Rashba spin-orbit coupling. The latter lifts the spin degeneracy and creates multiple spin-momentum locked Fermi surfaces. An external magnetic or Zeeman field is required to tune a quantum phase transition from an induced $s$-wave superconductor to a topological superconductor with Majorana fermions dispersing along the edges of 2D systems \cite{JaySau:prl10,JaySau:prb10,Alicea:prb10} or bounded to the end points of semiconductor nano-wires \cite{Lutchyn:prl10,Oreg:prl10}.       
 
At the heart of these settings for generating a topological phase with MBSs lies the existence of {\em three} conventional ingredients: a semiconductor quantum well with strong Rashba coupling, an $s$-wave superconductor, and a ferromagnetic insulator. The Zeeman field produced by the ferromagnetic layer should be strong enough to remove the extra Fermi surfaces, and simultaneously should be weak for the  induced superconducting pairing amplitude to survive, a condition which severely restricts the choice of materials. Moreover, the induced superconductting gap is rather small to allow for resolving MBSs \cite{Wang52}. Therefore, it is highly demanding to look for heterostructures with less impeding ingredients.  

The recently discovered superconductivity in epitaxial Bismuth/Nickel (Bi/Ni) bilayer heterostructure with relatively large transition temperature $T_{c}\approx 4.2$K may provide an example of an intrinsic topological superconductor with chiral $d_{xy}\pm i d_{x^2-y^2}$ order parameter \cite{Gong:sc17}. The nodeless structure of the proposed gap function is consistent with measurements of frequency-dependent optical conductivity in time-domain THz spectroscopy \cite{Chauhan:prl19}. It is also argued that the superconductivity could result from the bulk alloys near the interface \cite{Chao:prb19}, but it turns out the surface superconductivity is consistent with thinckness dependent of transition temperature \cite{Gong:sc17}. A schematic representation of the bilayer system in shown in Fig.~\ref{fig1}. The main advantage of the latter system over the quantum well structures discussed above is that here the superconductivity is intrinsically driven by the ferromagnetic fluctuations circumventing the proximity to an extra $s$-wave superconductor layer. Now the important question is: what are the vortex bound states in an intrinsic $d_{xy}\pm i d_{x^2-y^2}$ topological superconductor in Bi/Ni bilayer system? The aim of this paper is to theoretically answer this question. We show that the Berry phase effects change the angular momentum of the order parameter by one giving rise to odd parity with zero-energy states. Furthermore, in contrary to conventional superconductors, large ratios of superconducting gap to Fermi energy $\Delta/E_F\approx 10^{-2}-10^{-1}$ in Bi/Ni allow for MBS to remain well-separted from the low-lying excited states.   

For non-chiral $d$-wave superconductor $d_{x^2-y^2}$ the existence of exteneded \cite{Franz:prl98} and localized core states \cite{Knezevic:1999, Kato:2001} have been discucussed. The analoge of discrete Caroli-de Gennes-Matricon core states \cite{Caroli:pl64} for spin-rotatinally symetric chiral $d$-wave superconductors has been studied \cite{Franz:prl98, Rosenstein:13}, but no zero modes are reported. The heterostrucure of topological insulator Bi$_2$Se$_3$ film on a nodal $d$-wave superconductor Bi$_2$Sr$_2$CaCu$_2$O$_{8+\delta}$ has been experimentally studied recently where a rather large induced and isoptropic superconducting gap was reported \cite{Wang:nature2013}. The isotropically nodeless gap was attributed to an emergent s-wave component on the surface of TI due to broken four-fold symmetry with possible MBSs in vortex cores \cite{Li:prb15}. Note that our system is also distinct from the case of $d_{x^2-y^2}$ superconductor proximitized to an electron gas studied in Ref. [\onlinecite{Delgado:prb18}], where an external Zeeman field is applied to the system while in our work the time-reversal symmetry is already broken by the nature of chiral pairing.  
 
The paper is organized as follows. In Sec.~\ref{Sec:H0} we derive the single-particle Hamiltonian describing the helical electron states near the Fermi surface, then in Sec.~\ref{Sec:pairing} we derive the pairing correlations driven by magnetic fluctuations. In Sec.~\ref{Sec:MBS} we study the vortex bound states for various cases and the existence of zero-energy states, and finally Sec.~\ref{conclude} concludes.

\section{the real space projected non-interacting Hamiltonian\label{Sec:H0}}
 We begin with the electronic structure of the surface of Bi exposed to the vacuum as shown in Fig.~\ref{fig1}. Most likely, the interface adjacent to Ni doesn't contribute to superconductivity due to strong pair breaking effects, which is consistent with the thickness dependency of $T_c$ \cite{Gong:sc17,Gong:iopSc2015}. The strong Rashba coupling splits the surface electron states to spin-momentum locked states with the largest pocket centered around the center of surface of Brillouin zone. For simplicity we only consider this pocket and, hence, the Bi thin film in our setup in Fig.~\ref{fig1} can be replaced with a surface of TI as well. Therefore the helical electron states are described by the following action in Euclidean-time formalism: 

\be \label{Se} S_e= \int d\tau d\mathbf{r} \, \Psi^\dagger\left(\partial_\tau +\tilde{H}_{0}\right) \Psi. \ee 

The non-interacting Hamiltonian $\tilde{H}_{0}$ reads as 

\be\label{so_ham} \tilde{H}_{0}= \int d\mathbf{r}~\Psi^\dagger(\mathbf{r})\left(v_{F} [\boldsymbol{\sigma}\times\mathbf{p}]_z -\mu)\Psi(\mathbf{r}\right). \ee where $\Psi=(\psi_\uparrow , \psi_\downarrow)^T$ with $\psi_\updownarrow$ as electron annihilation operators, $v_F$ is the magnitude of spin-orbit interaction or equivalently the Fermi velocity of electrons at the surface of TI, $\mu$ is the chemical potential, $\boldsymbol{\sigma}$ is a vector of Pauli matrices, and $\mathbf{p}=-i\boldsymbol{\nabla}$. We set $\hbar=k_{B}=1$ throughout. 

Since we are interested in electron states near the Fermi surface, we use helical eigenstates $|\mathbf{k},\pm\ra = (1,\pm i e^{-i\phi_k})^T/\sqrt{2}$ in momentum space, where $\tilde{H}_{0}(\mathbf{k})|\mathbf{k},\pm \ra=\pm \varepsilon_{\mathbf{k}}|\mathbf{k},\pm \ra $ with $\varepsilon_{\mathbf{k}}=v_{F}|\mathbf{k}|$. We assume that the Fermi level crosses the ``+" band and is located well above the node. We first write the annihilation operators in the spin basis in terms of annihilation operators in the helical ``$\pm$" basis:
\be \label{projection} \psi_{\uparrow \mathbf{k}}=\frac{1}{\sqrt{2}}( \psi_{\mathbf{k}+} + \psi_{\mathbf{k}-}), \psi_{\downarrow \mathbf{k}}=\frac{i e^{-i\phi_\mathbf{k}}}{\sqrt{2}}( \psi_{\mathbf{k}+}  -  \psi_{\mathbf{k}-}). \ee 

Then in real space they become 
\bea &&\psi_\uparrow(\mathbf{r})= \frac{1}{\sqrt{2}}[\psi_{+} (\mathbf{r}) + \psi_{-} (\mathbf{r})],\\ \label{psidown}
 &&\psi_\downarrow(\mathbf{r})=\frac{i}{\sqrt{2}}\sum_\mathbf{k} e^{-i\phi_\mathbf{k}}(\psi_{\mathbf{k}+}-\psi_{\mathbf{k}-})e^{i\mathbf{k}\cdot\mathbf{r}}. \eea
 
 To perform the momentum sum in Eq. (\ref{psidown}) we approximate the angular exponential term using
 \be \label{phase_approx} e^{-i\phi_\mathbf{k}}\approx \frac{k_x}{k_F}-i \frac{k_y}{k_F}, \ee
 which is justified so long as the electron states near the Fermi surface are involved in the physical processes of interest such as pairing instabilities. Here $k_{F}=\mu/v_{F}$ is the Fermi momentum. The field operator $\psi_\downarrow(\mathbf{r})$ then reads as  
\be \psi_\downarrow(\mathbf{r})= \frac{1}{\sqrt{2}k_{F}} (\partial_x-i\partial_y)\left[\psi_{+}(\mathbf{r})-\psi_{-}(\mathbf{r})\right]. \ee

By projection to the Fermi surface we obtain 
\be \psi_\uparrow(\mathbf{r})\approx\frac{1}{\sqrt{2}}\psi_{+}(\mathbf{r}), \, \psi_\downarrow (\mathbf{r}) \approx \frac{1}{\sqrt{2}k_F}(\partial_x - i\partial_y) \psi_{+}(\mathbf{r}). \ee

Rewritting Eq. (\ref{so_ham}) by use of these expressions, the projected form of the non-interacting Hamiltonian reads as follows: 
\be\label{H0} H_{0}= - \int d\mathbf{r} \, \psi_{+}^\dagger(\mathbf{r})\left( \frac{v_F}{k_F}\nabla^2+\mu\right)\psi_{+}(\mathbf{r}),\ee
which is not dissimilar to the Hamiltonian of a 2D Fermi gas via the identification $v_F/k_{F}=1/2m$. Hereafter we drop the subindex in the field and write $\psi_{+}(\mathbf{r})\equiv\psi(\mathbf{r})$.

\section{Model of magnetic fluctuations and pairing hamiltonian \label{Sec:pairing}}
 To make the structure of the paper self-contained, in this section we present the details of a minimal model describing the superconductivity in the bilayer structure shown in Fig.~\ref{fig1} which is largely based on the Ref. [\onlinecite{Gong:sc17}]. In the regime of interest the superconducting $T_{c}$ is much lower than the Curie temperature of ferromagnetic Ni layer, and hence the ferromagnet is deep inside the ordered phase. We assume that the in-plane moments are aligned along the $y$ direction (see Fig.~\ref{fig1}) and the low-energy fluctuations of the magnetic moments, the spin waves, are described by the vector $\mathbf{l}(\tau,\mathbf{r})$ in which $\mathbf{l}\cdot \hat{y}=0$. The magnetic fluctuations and their coupling to electrons are described, respectively, by the following actions\cite{Kargarian:prl16}:    
 
 \be \label{SM} S_M= \frac{\rho_s}{2} \int d\tau d\mathbf{r} \, \left[ -i(\mathbf{l}\times \partial_\tau \mathbf{l})+ \kappa (\nabla \mathbf{l})^2 \right], \ee
 and 
\be \label{SeM} S_{eM}= g \int d\tau d\mathbf{r} \, \Psi^\dagger \left(\mathbf{l}\cdot\boldsymbol{\sigma}\right) \Psi, \ee 
where $\rho_s$ is the density of magnetic moments, $\kappa$ is the characteristic of spin waves, and $g$ is the strength of interaction between electron spins and magnetic moments. For simplicity and in the interest of formation of Cooper pairs with zero center-of-mass momentum we only consider the out-of-plane fluctuations denoted by $l_z\equiv b$.  

By taking Fourier transform to momentum space, the Eqs. (\ref{SM}-\ref{SeM}) become  
\be S_M=\frac{T}{2} \sum_q D^{-1}(\mathbf{q})b^\dagger_q b_q, \ee and 
\be S_{eM}=\frac{gT}{2}\sum_{k,q,\alpha\beta}\left( b_q \psi^\dagger_{k\alpha} \sigma^z_{\alpha\beta} \psi_{k+q,\beta} + \mathrm{h.c.}\right),\ee
where $T$ is the temperature, $q=(\mathbf{q},2n\pi T)$, $k=(\mathbf{k},(2n+1)\pi T)$ with $n$ as an integer, and $D(\mathbf{q})=1/(\kappa \rho_s |\mathbf{q}|^2+\zeta)$ is the magnon propagator with a small gap $\zeta$ due to anisotropy\cite{Tserkov:prl12}. Upon integrating out the bosonic field $b$ and subsequent projection of electron fields $\psi_{k\alpha}$ to the Fermi surface described by effective spinless fermion operators $\psi_{k}$ in Eq. (\ref{projection}), we obtain the following effective interaction between electrons in the Cooper channel\cite{Gong:sc17}  

\be \label{Sc} S_{c}=\frac{T}{2\mathcal{A}}\sum_{k,k'} U(\mathbf{k},\mathbf{k}')e^{-i(\phi_k-\phi_{k'})} \psi^\dagger_k  \psi^\dagger_{-k}  \psi_{-k'} \psi_{k'}, \ee where $\mathcal{A}$ is the area of the system. Here $U(\mathbf{k},\mathbf{k}')$ is an even function of its arguments and can be expanded in angular harmonics as
\be \label{U} U(\mathbf{k},\mathbf{k}')=\sum_{l=\textit{even}} U_l ~ e^{il(\phi_k-\phi_{k'})}. \ee 

Therefore the even angular momentum components of the interaction matrix contribute to the odd component of the condensate $f=\langle  \psi_{-k} \psi_{k} \rangle$. This is a direct result of the non-trivial topology of Dirac Fermions. The effective angular momentum of the condensate $f$ is decreased by one due to the Berry phase. Inserting Eq. (\ref{U}) in  Eq. (\ref{Sc}) and decoupling the interaction in the Cooper channels $f$, we obtain the mean-field BCS Hamiltonian as follows.

\be\label{pairing} H_\Delta=\sum_{\mathbf{k}} \Delta(|\mathbf{k}|)e^{i(l-1)\phi_k}  \psi^\dagger_k  \psi^\dagger_{-k} \,\, + \,\, \mathrm{h.c.}\ee

In this work we only consider the channels with the lowest angular momenta $l=0,\pm 2$. The superconducting instability in channels $l=\pm2$ is driven by the magnetic fluctuations being relevant to Bi/Ni bilayer system, while the instability with $l=0$ arises from phonons or charge fluctuations\cite{Gorkov:prl01}, i.e., $\sigma^{z}\rightarrow \mathbf{1}$ in Eq. (\ref{SeM}). In our formalism below we study all cases.

\section{spectrum of vortex bound states \label{Sec:MBS}}
The formulation and arguments presented in preceding sections provide a minimal superconducting Hamiltonian using Eqs. (\ref{H0}) and (\ref{pairing}), i.e., $H=H_{0}+H_{\Delta}$. In the following subsections we first derive the corresponding Bogoliubov-de Gennes (BdG) equations for each channel $l$ and then study the spectrum of vortex bound states.    

\subsection{The cases with $l=2$ and $l=0$}
By inspection we see that for both cases the phases of the pairing in Eq.(\ref{pairing}) are simply complex conjugate of each other, and thus, they can be treated within the same formalism. We present the details of calculations for $l=2$ and will briefly discuss the $l=0$ case at the end of this subsection. For the former case the pairing term $H_{\Delta}$ in Eq. (\ref{pairing}) is written as 

\be H_\Delta=\sum_{\mathbf{k}}\frac{\Delta(|\mathbf{k}|)}{k_F}(k_x+ik_y)\psi^\dagger_k \psi^\dagger_{-k} \,\, + \,\, \mathrm{h.c.}, \ee
where we use Eq. (\ref{phase_approx}), assuming pairing occurs near the Fermi surface. To introduce a vortex in the order parameter, we assume that the space profile of pairing gap in the polar coordinate is $\Delta(\mathbf{r})=\Delta(r)e^{in\theta}$, where $r$ is measured from the center of vortex and $n$ denotes the winding of the vortex, the degree of vorticity. Thus, the full mean-field Hamiltonian of this system in real space can be recast as 

\begin{multline}\label{h_mf}
H=\int \, d\mathbf{r}\, \left[ - \psi^\dagger \left(\frac{v_F}{k_F}\nabla^2 +\mu\right)\psi + i \frac{\Delta(r)}{2k_F}\psi^\dagger \{e^{in\theta},\partial_x \right. \\ \left. +i\partial_y \}\psi^\dagger + i\frac{\Delta(r)}{2k_F}\psi \{e^{in\theta},\partial_x-i\partial_y \}\psi \right],
\end{multline}
where $\{A,B\}=(AB+BA)/2$ is a symmetric operator. We define $\gamma_i^\dagger$ as a creation operator for the $i$-th eigenstate of the mean-field Hamiltonian satisfying 
\be\label{gamma_com}[H_{MF},\gamma_i^\dagger]=E_i \gamma_i^\dagger.\ee

The operator $\gamma^\dagger$ is used to diagonalize the Hamiltonian and hence can be written as a linear combinations of $\psi$'s:

\be\label{lin_com} \gamma_i^\dagger=\int d\mathbf{r}\, (u_i(\mathbf{r})\psi^\dagger(\mathbf{r})+v_i(\mathbf{r})\psi(\mathbf{r})). \ee

From now on we drop the index $i$ for simplicity. Using Eq. (\ref{h_mf}) and Eq. (\ref{lin_com}) in Eq. (\ref{gamma_com}), we get a system of differential equations of the form $H_{BdG}\varphi(\mathbf{r})=E\varphi(\mathbf{r})$, where $\varphi(\mathbf{r})=(u(\mathbf{r}),v(\mathbf{r}))^T$ and $H_{BdG}$ is the BdG Hamiltonian 

\begin{small}
\be\label{h_bdg} H_{BdG}=\bpm -\frac{v_F}{k_F}\nabla^2-\mu && i\frac{\Delta(r)}{2k_F}\left\{ e^{in\theta},\partial_x+i\partial_y \right\} \\ i\frac{\Delta(r)}{2k_F}\left\{ e^{-in\theta},\partial_x-i\partial_y \right\} && \frac{v_F}{k_F}\nabla^2+\mu \epm. \ee
\end{small}

Rewriting the differential operators in polar coordinates, Eq. (\ref{h_bdg}) assumes the following form:
\begin{widetext}
\be\label{h_bdg_cyn} H_{BdG}=\bpm -\frac{v_F}{k_F}\left[\frac1r \partial_r(r\partial_r)+\frac{1}{r^2}\partial^2_\theta \right] -\mu & i \frac{\Delta(r)}{k_F}e^{in'\theta}\left(\partial_r + i \frac{1}{r}\partial_\theta-\frac{n}{2r}\right) \\ \\ i \frac{\Delta(r)}{k_F}e^{-in'\theta}\left(\partial_r -i \frac{1}{r}\partial_\theta-\frac{n}{2r}\right) & \frac{v_F}{k_F}\left[\frac1r \partial_r(r\partial_r)+\frac{1}{r^2}\partial^2_\theta \right]+\mu \epm, \ee 
\end{widetext}
where $n'=n+1$. We use a pseudo-rotation operator defined by the unitary transformation $U(\theta)=e^{-i(m+\frac{n'}{2}\tau^z)\theta}$, where $\tau^z$ is the Pauli matrix acting in particle-hole space, to remove the phase dependency of the pairing gap \cite{JaySau:prb10}. That is, we write the wave function $\varphi(\mathbf{r})$ in the form $\varphi(r,\theta)=e^{i(m+\frac{n'}{2}\tau^z)\theta}\varphi(r)$. The possible values for $m$ are determined by the single-valued condition of wave function implying that $m$ is an integer (half-integer) for even (odd) $n'$. Using this transformation the eigenvalue problem turns into a set of differential equations for $v(r)$ and $u(r)$ as 
\begin{widetext}
\bea \label{bdg_1} 
&&\partial^2_r u + \frac1r \partial_r u - \frac{m_+^2}{r^2}u+\frac{k_F \mu}{v_F} u - i \frac{\Delta(r)}{v_F}\left(\partial_r v - \frac{2m-1}{2r}v\right)=-\frac{k_F E}{v_F}u, \\ \label{bdg_2} 
&&\partial^2_r v + \frac1r \partial_r v - \frac{m_-^2}{r^2}v+\frac{k_F \mu}{v_F} v + i \frac{\Delta(r)}{v_F}\left(\partial_r u + \frac{2m+1}{2r}u\right)=\frac{k_F E}{v_F}v,
\eea
\end{widetext}
where $m_\pm = (2m \pm n')/2$. We see that the equations are not symmetric under $n \rightarrow -n$. Note that the shift in $n'$ relative to winding $n$ by one comes from the $\pi$-Berry phase of the electron states on the Fermi surface. The latter phase shifts the relative angular momentum of pairs by one \cite{Zeyt:18}. Therefore the bound states of cores with opposite phase winding around the vortex would have different energy spectra.  A set of equations similar to those quoted in Eqs.(\ref{bdg_1}-\ref{bdg_2}) is presented for a $p$-wave superconductor\cite{Read:prb2000, Matsumoto:prb01}, where the kinetic terms are usually in the long wavelength limit and it's assumed that the chemical potential is negative in the vortex core (the strong coupling phase) and positive outside (the weak coupling phase) with $\Delta(r)$ as a constant. In our case however we assume that $\mu$ to be constant and take a space-varying order parameter like conventional superconductors. 

Due to the $p$-wave structure of the Hamiltonian (\ref{h_bdg_cyn}) the vortex core hosts a Majorana fermion. In Appendix \ref{AppendixA} we employ an approach similar to Ref.[\onlinecite{JaySau:prb10}] and explicitly show that a zero-energy state exists in the vortex core. Furthermore, in order to get more insight into the spectrum of the bound states we use a long-wave approximation\cite{Caroli:pl64,Kopnin:prb91} for the wave function (see Appendix \ref{AppendixB} for details) and show that the spectrum reads as 
\begin{equation}\label{spec}
\tilde{E}= m \, \frac{\int_0^\infty (\frac{\tilde{\Delta}(x')}{x'}+ \frac{n'}{x'^2}) e^{-2\chi(x')} \, dx'}{\int_0^\infty e^{-2\chi(x')}\, dx'}
\end{equation} 
where $\chi(x)=\int_0^x \frac{\tilde{\Delta}(x')}{2}\, dx'$, and we used dimensionless parameters $x=k_{F}r$ and $\tilde{\Delta}=\Delta/\mu$ and $\tilde{E}=E/\mu$. One has to note that we are a little cavalier in using the semiclassical approach, since the value of $\Delta/E_F$ in our system, as we discuss more in Sec.\ref{conclude}, is rather large compared to conventional superconductors. Within the approximations used it turns out that the vortices with $n'\neq 0$ would have very large energy if $m\neq 0$ simultaneously. Let us consider a vortex with the lowest value of vorticity $n'=0$ corresponding to $n=-1$ as shown schematically in Fig. \ref{fig1}. The condition $U(2\pi)=1$ implies that the $m$ has to be an integer number with $m=0$ corresponding to a zero-energy state.  

As we mentioned at the beginning of this subsection the cases with $l=2$ and $l=0$ can be treated on equal footing, since the corresponding equations are the same. The latter case, $l=0$, yields $n'=n-1$ and a vortex with lowest winding number will have $n=1$. Again the vortex can host a zero-energy state.

\subsection{The Case with $l=-2$}

In this case the BdG equations become third order and an analytical solution for them is a formidable task if not impossible. To circumvent this problem, we use the semi-classical approximation used in the analysis of the Andreev bound states in superconductors\cite{Andreev:jetp64} and closely follow Refs.[\onlinecite{Stone:prb96}] and [\onlinecite{Volovik:99}]. The BdG equation in this case is:

\begin{multline} \label{bdg_-2}
H_{BdG}= (h_{0}-\mu)\tau_3 +  i\frac{\Delta(r)}{2k_F^3}\left\{ e^{in\theta},(\partial_x+i\partial_y)^3 \right\}\tau_+ \\ + i\frac{\Delta(r)}{2k_F^3}\left\{ e^{-in\theta},(\partial_x-i\partial_y)^3 \right\}\tau_{-}, 
\end{multline} where $h_{0}=-\frac{v_F}{k_F}\nabla^2$ is the kinetic energy.
For solving BdG equations we use an ansatz for the wave function as $\Psi=\varphi(\mathbf{r}) e^{i\mathbf{q}\cdot\mathbf{r}}$ with an approximation that the momentum $\mathbf{q}$ is restricted to the Fermi surface, i.e. $\mathbf{q}=k_F(\cos \phi , \sin \phi)$ known as the momentum of a quasi-particle in the Andreev approximation. Using $\Psi$ in Eq. (\ref{bdg_-2}), we obtain 
\begin{multline} \label{bdg_andreev}
H=-i\mathbf{v}\cdot\nabla \tau_3 + \Delta(r) \cos(\theta')\tau_1+\Delta(r)\sin(\theta')\tau_2,  \end{multline} which acts only on $\varphi(\mathbf{r})$. Here we defined $\mathbf{v}=(2v_{F}/k_{F})\mathbf{q}$ and $\theta'=n\theta+3\phi$. We rotate the coordinates such that the new $x$-axis becomes parallel to $\mathbf{q}$:
\begin{multline} H=-iv \partial_x \tau_3+\Delta(r) \cos(n\theta+(3-n)\phi)\tau_1 \\ +\Delta(r)\sin(n\theta+(3-n)\phi)\tau_2 \end{multline}
Then the $\phi$ dependence in the Hamiltonian can be removed using the transformation $\varphi \rightarrow e^{i(3-n)\phi\tau_3/2} \varphi$: 
\be \label{quasi1D} H=-iv \partial_x \tau_3 + \Delta(r) \cos(n\theta)\tau_1+\Delta(r)\sin(n\theta)\tau_2. \ee

This is a quasi one-dimensional problem derived in Ref.[\onlinecite{Stone:prb96}] (see Eq. (3.10) in the latter reference with replacement $\theta\rightarrow -n\theta$). To proceed we define an impact parameter for quasi-particles as $b=r\sin \theta$ which measures the minimum distance of the quasi particle trajectory from the origin of the vortex core. For the pairing gap we use a profile as $\Delta(r)=\Delta\Theta(r-R)$, where $\Theta(x)$ is the usual step function and $R$ is the radius of the vortex. The latter is of order of $R\simeq v_{F}/\Delta$. With these assumptions the quasi one-dimensional model Eq.(\ref{quasi1D}) can be solved to obtain the energy spectrum of the bound states. For small values of $bk_{F}\ll1$, corresponding to trajectories passing near the origin, the spectrum reads as 

\bea E_{m}=\omega_{0}\left(-n\pi+2\pi\left(m+\frac{1}{2}\right) \right),\eea
where $\omega_{0}=v_{F}/2R$ is the angular velocity of the superfluid \cite{Stone:prb96}. Now it is clearly seen that for $n=1$ the spectrum becomes $E_{m}=2\pi\omega_{0}m$ and a zero mode corresponds to $m=0$. Therefore the vortex bound states for the $l=-2$ case also contain a zero-energy mode.



\section{Conclusions\label{conclude}}
This work is mainly motivated by the efforts put forward in recent years to find Majorana bound states in the vortex core of  superconducting states. We proposed the superconducting epitaxial Bi/Ni bilayer as a platform to create and manipulate the Majorana states. The system has an advantage over the heterostructues proposed in the literatures in that the chiral superconducting states are created intrinsically due to the magnetic fluctuations of the ferromagnetic layer circumventing the need for a proximixed superconducting layer. The heterostructure here can be replaced by other materials combinations, e. g., a thin film of topological insulator Bi$_2$Te$_{3}$ deposited on the magnetic insulator layer FeTe \cite{Manna:nature2017,He:nature2014}, or superconducting states in oxide interfaces \cite{Scheurer:nature2015}, making our proposal for creating and manipulating of zero modes experimentally feasible. 

The chiral superconducting states in Bi/Ni bilayer are characterized by total angular momentum $l=\pm2$ corresponding to $d_{xy}\pm d_{x^2-y^2}$, which break the time-reversal symmetry. We showed that the underlying strong spin-orbit coupling alter the bound state spectrum in the vortex core. In particular we demonstrated that a zero-energy state corresponding to Majorana bound state appears at the vortex core for both cases of the pairing wave functions. We also showed that the case with total angular momentum $l=0$, which intrinsically does not break the time-reversal symmetry and my be induced by charge fluctuations, can also support a zero-energy state. The set-up studied in our work, as shown in Fig.~\ref{fig1}, should be contrasted with proposals in the literatures where the superconductivity is induced by proximity. Our finings may motivate the search for Majorana zero modes in vortices in heterostructures with no need for proximity to an extra superconducting layer.

Another peculiar aspect of vortex bound states in Bi/Ni is that the Majorana bound state remains well separated from the low-lying excited states due to a relatively large value of $\Delta/E_F$. The superconducting gap is estimated to be about $\Delta\approx 0.7$~meV \cite{Chauhan:prl19} and the Fermi energy $E_F$ is about 27~meV for hole pockets and 10~meV for electron pockets \cite{Hofmann2006} yielding a ratio of about $10^{-2}-10^{-1}$. The ratio is by an order of magnitude larger than the corresponding values for conventional superconductors. A clear observation of discrete bound states in iron-based superconductor FeTe$_{0.55}$Se$_{0.45}$\cite{Chen:NCOM2018} is reported due to large value of $\Delta/E_F$. The surface of latter compound was shown to be a topological superconductor \cite{Zhang:Sci18} hosting a well-resolved Majorana bound state \cite{Wang:Sci18}. Therefore the same sort of well-resolved bound states and Majorana zero mode should be observed in epitaxial Bi/Ni bilayer.          

In summary, our work offers the chiral superconductor in epitaxial Bi/Ni bilayer as a new platform to explore the vortex states with (i) robust Majorana bound state due to large ratio of $\Delta/E_F$, (ii) less complexity in the heterostructure, and (iii) relatively high transition temperature.

\section{Acknowledgments} 
The authors would like to acknowledge the support from the Sharif University
of Technology under Grant No. G690208.


\appendix 
\section{Explicit calculation of zero energy state \label{AppendixA}}
Using the dimensionless parameters $x=k_{F}r$ and $\tilde{\Delta}=\Delta/\mu$ and $\tilde{E}=E/\mu$, the Eqs.(\ref{bdg_1}-\ref{bdg_2}) can be rewritten as 

\begin{widetext}
\bea \label{i}
\partial^2_x u + \frac1x \partial_x u -\frac{m_+^2}{x^2}u +u - i \tilde{\Delta}(x)\left(\partial_x v - \frac{2m-1}{2x}v\right)=-\tilde{E}u\\
\label{ii}
\partial^2_x v + \frac1x \partial_x v -\frac{m_-^2}{x^2}v +v + i \tilde{\Delta}(x)\left(\partial_x u + \frac{2m+1}{2x}u\right)=\tilde{E}v.
\eea
\end{widetext}

Here we consider (\ref{i}) and (\ref{ii}) and follow Ref.[\onlinecite{JaySau:prb10}] for $\tilde{E}=0$. We assume that the vortex boundary is at $x=x_0$ and take $\tilde{\Delta(x)}$ to be zero for $x<x_0$ and a constant value for $x>x_0$. For the region inside the vortex the equations become a set of decoupled Bessel equations with the general analytical solution:
\begin{equation}
\bpm u(x) \\ v(x) \epm= \bpm A_1 J_{m_+}(x) \\ A_2 J_{m_-}(x) \epm
\end{equation}
Where $J_m(x)$ is the Bessel function of the first kind. A closed solution of equations for $x>x_0$ is not tractable. Instead we try to find the asymptotic solution of equations in the limit $x\gg 1$. In this limit (\ref{i}) and (\ref{ii}) become:
\begin{equation}
\partial^2_x u + u - i \tilde{\Delta}\partial_x v=0
\end{equation}
\begin{equation}
\partial^2_x v + v + i \tilde{\Delta}\partial_x u=0
\end{equation}
Looking for a decaying solution of the form $\bpm u \\ v \epm = \bpm u_0 \\ v_0 \epm e^{\kappa x}$ one gets:
\begin{equation}
\bpm u(x) \\ v(x) \epm = A_3 \bpm 1 \\ -i \epm f_+(x) + A_4 \bpm 1 \\ -i \epm f_-(x)
\end{equation}
Where $f_\pm(x)$ are decaying functions with the asymptotic form $f_\pm(x) \rightarrow e^{-\kappa_\pm x}$ and $\kappa_\pm = \frac{|\tilde{\Delta}|}{2}\pm \sqrt{(\frac{\tilde{\Delta}}{2})^2-1}$. We have to match the solutions for inside and outside of the vortex at the vortex boundary. The condition $\varphi(x_0^-)=\varphi(x_0^+)$ gives:
\begin{equation}\label{u}
A_1 J_{m_+}(x_0)=A_3 f_+(x_0) + A_4 f_-(x_0)
\end{equation}
\begin{equation}\label{v}
A_2 J_{m_-}(x_0)=-i(A_3 f_+(x_0) + A_4 f_-(x_0))
\end{equation}
$\varphi'(x_0^-)=\varphi'(x_0^+)$ yields:
\begin{equation}\label{u'}
A_1 J'_{m_+}(x_0)=A_3 f'_+(x_0) + A_4 f'_-(x_0)
\end{equation}
\begin{equation}\label{v'}
A_2 J'_{m_-}(x_0)=-i(A_3 f'_+(x_0) + A_4 f'_-(x_0))
\end{equation}
These equations alongside with the normalization condition $\int (u^2(r)+v^2(r))d\mathbf{r}=1$ should be satisfied in order to have a solution. In general, it is not possible to satisfy all of these conditions by adjusting only 4 unknowns ($A_1$-$A_4$) and the problem is over-constrained and a zero mode solution does not exist. Nevertheless, in the special case of $n'=0$ where $m_+=m_-$ from (\ref{u}) and (\ref{v}) we have $A_2=-iA_1$ which makes (\ref{u'}) and (\ref{v'}) identical and therefore there are only 3 independent boundary conditions which alongside with the normalization condition assign a unique value to $A_1$-$A_4$. This is compatible with (\ref{spec}) in which a zero energy state exists only if $n'=0$.\\ 

\section{Energy spectrum of bound states \label{AppendixB}}

Following Ref.[\onlinecite{Kopnin:prb91}] and Ref.[\onlinecite{Caroli:pl64}], we assume that the wave functions in (\ref{i}) and (\ref{ii}) take the following form:
\begin{equation}
\begin{pmatrix} u \\ v \end{pmatrix} = \begin{pmatrix} f_{+} \\ g_{+} \end{pmatrix} H^{1}_q(x) + \begin{pmatrix} f_{-} \\ g_{-} \end{pmatrix} H^{2}_q(x), 
\end{equation}
where $H^{1}_q$ and $H^{2}_q$ are the Hankel functions of first and second kinds, respectively, and $f$ and $g$ are slowly varying functions. We insert the above ansatz in Eqs. (\ref{i}-\ref{ii}) and neglect the second derivatives of $f_{\pm}$ and $g_{\pm}$. Using the asymptotic behavior of Hankel functions, we obtain the following differential equations governing $f_{\pm}$ and $g_{\pm}$: 
 \begin{equation}
\frac{df_{\pm}}{dx}-i\frac{\tilde{\Delta}}{2}g_{\pm}=\pm i \left(\frac{\tilde{E}}{2}- \frac{n'm}{2x^2}\right)f_{\pm}\mp \frac{\tilde{\Delta}m}{2x}g_{\pm}
\end{equation}
\begin{equation}
\frac{dg_{\pm}}{dx}+i\frac{\tilde{\Delta}}{2}f_{\pm}=\mp i\left(\frac{\tilde{E}}{2}- \frac{n'm}{2x^2}\right)g_{\pm} \mp \frac{\tilde{\Delta}m}{2x}f_{\pm}.
\end{equation}

The low-energy spectrum of bound states in the vortex lies within the superconducting bulk gap. Therefore a natural assumption is to assume $\tilde{E}<\tilde{\Delta}$ in the equations above, otherwise there would be no bound states at the vortex core. Physically the bound states result from the Andreev reflections of quasiparticles in the vortex core \cite{Stone:prb96}. We also assume that $x\gg 1$ which means that we are considering the long distance behavior of the system. Then by treating the expressions on the right-hand side as perturbations, we obtain the following expressions for $f$ and $g$ up to first order:
\begin{widetext}
\bea 
\begin{pmatrix} f_1 \\ g_1 \end{pmatrix}=A_1 \left\{ \begin{pmatrix}1 \\ i \end{pmatrix}e^{-\chi(x)} \right. \left. -  \begin{pmatrix}
i \\ 1
\end{pmatrix}e^{\chi(x)} \int_x^\infty \left(\frac{\tilde{\Delta}m}{2x'}+ \frac{n'm}{2x'^2} - \frac{\tilde{E}}{2}\right)e^{-2\chi(x')} \, dx' \right\}\\
\begin{pmatrix} f_2 \\ g_2 \end{pmatrix}=A_2 \left\{ \begin{pmatrix}1 \\ i \end{pmatrix}e^{-\chi(x)} +  \begin{pmatrix}
i \\ 1
\end{pmatrix}e^{\chi(x)} \int_x^\infty \left(\frac{\tilde{\Delta}m}{2x'} + \frac{n'm}{2x'^2} - \frac{\tilde{E}}{2}\right)e^{-2\chi(x')} \, dx' \right\},
\eea
\end{widetext}
where $\chi(x)=\int_0^x \frac{\tilde{\Delta}(x')}{2}\, dx'$. In order to avoid the singularity of Hankel functions at the origin we should take $A_1=A_2$ and the second terms should vanish as $x\rightarrow 0$.  These boundary conditions eventually lead to the following expression for the energy spectrum of vortex bound states  $\tilde{E}$ in (\ref{spec}).

%

\end{document}